\documentclass[12pt]{iopart}

\usepackage{cite}
\usepackage[utf8]{inputenc}
\usepackage{graphicx}
\usepackage{graphicx,epstopdf}
\usepackage{dcolumn}
\usepackage{amsmath}
\usepackage{lineno,hyperref}
\usepackage{amsfonts}
\usepackage{amsthm}
\usepackage{bbm}
\usepackage{amssymb}
\usepackage{mathrsfs}
\PassOptionsToPackage{caption=false}{subfig}
\usepackage{subfig}
\usepackage{color}
\usepackage[normalem]{ulem}

\newcommand{\Pcal}{\mathcal{P}}

\newcommand{\Ucal}{\mathcal{U}}

\usepackage{blindtext}
\usepackage{lipsum}
\usepackage{bbm}

\newcommand{\1}{\mathbbm{1}}

\newcommand{\ket}[1]{| #1 \rangle}

\newcommand{\interpro}[2]{\langle #1 | #2 \rangle}
\newcommand{\bra}[1]{\langle #1 |}
\newcommand{\trs}[1]{ \text{Tr}[ #1 ]}

\usepackage{etoolbox}

\makeatletter
\newcommand{\mainmatter}{%
	\setcounter{footnote}{0}%
	\patchcmd{\@makefntext}{\fnsymbol}{\arabic}{}{}%
	\patchcmd{\@thefnmark}{\fnsymbol}{\arabic}{}{}%
	\def\@makefnmark{\textsuperscript{\arabic{footnote}}}%
}
\makeatother

\begin{document}

\title[Non-Markovian effects on charging and self-discharging processes of quantum batteries]{Non-Markovian effects on charging and self-discharging process of quantum batteries}

\author{F. H. Kamin$^1$, F. T. Tabesh$^{1,\ast}$, S. Salimi$^{1,\dagger}$, F. Kheirandish$^1$, Alan C. Santos$^{2}$}

\address{$^1$Department of Physics, University of Kurdistan, P.O.Box 66177-15175, Sanandaj, Iran.\\$^2$Departamento de Física, Universidade Federal de São Carlos, Rodovia Washington Luís, km 235 - SP-310, 13565-905 São Carlos, SP, Brazil.}
\ead{$^\ast$shsalimi@uok.ac.ir}
\vspace{10pt}
\begin{indented}
\item[]February 2014
\end{indented}

\begin{abstract}
The performance of quantum technologies that use entanglement and coherence as resource is highly limited by decohering effects due to their interaction with some environment. Particularly, it is important to take into account situations where such devices unavoidably interact with a surrounding. Here, we study memory effects on energy and ergotropy of quantum batteries in the framework of open system dynamics, where the battery and charger are individually allowed to access a bosonic environment. Our investigation shows that
the battery can be fully charged and its energy can be preserved for long times in non-Markovian dynamics compared with Markovian dynamics. In addition, the total stored energy can be completely extracted as work and discharge time becomes more longer as non-Markovianity increases. Our results indicate that memory effects can play a significant role in improving the performance of quantum batteries.
\end{abstract}

\section{Introduction}

Quantum batteries (QBs) are quantum systems that are capable to store energy, making it available for later goals. In fact, well known examples of classical batteries are electrochemical devices converting chemical energy to electrical one. They can be disposable or rechargeable using electricity and so on. Therefore, batteries are very comfortable for their multiple use, and their presence in everyday life has transformed them to essential elements. The role of quantum effects on the problem of energy storage has been extensively studied in last years~\cite{Alicki:13,Binder:15,PRL2017Binder,Ferraro:18,Le:18,Andolina:18,PRB2019Batteries,PRL_Andolina}. On the other hand, studies over the past few decades show that many modern-day technologies are based on the principles of quantum mechanics~\cite{Giovannetti:03,PRL2013Huber,DeffnerReview} can be more efficient than their classical ones. Then, we believe that quantum mechanical effects can be useful for miniaturized batteries working at quantum realm. In recent years, the study of QBs have attracted much more attention (there is worldwide interest in exploiting QBs) as small quantum systems used for temporary energy storage and transferring it from production centers to consumption centers \cite{Alicki:13}.

As mentioned above, the research on QBs and the investigation of  stability of stored energy are very important. Recently, there have been many efforts to design protocols whose main purpose is to control charging or discharging processes (such as ability to conserve energy and the stability of quantum batteries ) in order to achieve more operational quantum batteries~\cite{Santos:19-a,Santos:19-c,Andolina:19-2,Pirmoradian:19}. In most research, QBs are considered as closed systems, that do not interact with their environments. However, quantum systems are open systems where they undoubtedly interact with their surroundings. Also, open system dynamics can be characterized by either Markovian or non-Markovian approach~\cite{Petruccione:Book,Breuer:16}. In general, Markovian approximation is found on the suppositions that the coupling between the system and its environment is weak and  the initial total state is factorized. If any one of these conditions has been ignored, it may lead to non-Markovian dynamics. Existence of memory effects in evolution is insured by non-Markovian dynamics~\cite{Fanchini:14,Haseli:14,Luo:12,Breuer:09,Wolf:08,Rivas:10,Garcapintos:19,Obando:15,Passos:19-b,Paula:16,Lorenzo:13}. Since the charging process of QBs has been considered in Markovian regime~\cite{Garcapintos:19,PRB2019Batteries}, the case where non-Markovian effects are present is yet a point to be addressed. Hence, it is worthwhile investigating the quantum memory effects in order to counteract the energy dissipation in the process of charging or discharging and the maximum extractable work from QBs in such scenarios

In this paper, we study the role of dissipation and back-flow of information on the energy, ergotropy and discharge time of QB. For
this purpose, we consider a battery $B$ connected to a charger $A$, each of which interacts separately with its environment. The battery and the charger are two-level systems. We accurately obtain the time evolution of the total system and then examine the charging process efficiency (power) and the amount of work that can be extracted from the QB for both Markovian and non-Markovian dynamics. Specifically, we study the different ways in which the dissipation environment and coupling with the charger contribute to the
process. In particular, the non-Markovian dynamics plays a substantial role in the underdamped regime by increasing the energy transfer power from the charger to the battery. Moreover, non-Markovian dynamics may provide a charging performance near to the ideal case (no decohering effects). To end, we study an important phenomenon called \textit{self-discharging}, which limits the performance of QBs in storing the energy for long times~\cite{Santos:19-a}. We show how non-Markovian dynamics and its consequent memory effects allow for increasing the self-discharge time of the battery.

\section{Quantum batteries}

In this section, we briefly introduce the system we will use throughout this paper.
A QB is a physical quantum system that stores energy, where the advances in stored charge and charging process of such devices are associated with quantum mechanical phenomena. For this reason, it is not possible to say that any system with discrete energy spectrum can therefore be considered as a QB. The performance of QB does not emerge from discrete spectrum, but the usage of quantum correlations and/or coherence~\cite{Santos:19-c}. In order to define the energy scale of the QB we start by considering a time-independent reference Hamiltonian $H$, where the Hilbert space of the battery is assumed to be finite. In general, the available energy that can be extracted from a QB is not given by the difference of internal energy of the system. Such amount of energy is well defined from the \textit{ergotropy} of the system given by
\begin{align}
\mathcal{W} = \tr(\rho H) - \max_{U \in \Ucal} \tr([U\rho U^{\dagger}]H) \text{ , } \label{a1}
\end{align}
where $\rho$ is the initial system state and we maximize over the set of all accessible unitary operations $\Ucal$. In particular, in cases where $\rho$ is a pure state we can show that the optimal $U$ is so that $U\rho U^{\dagger}\!=\!\rho_{\text{gs}}$, where $\rho_{\text{gs}}$ is the ground state of the reference Hamiltonian $H$~\cite{Santos:19-a,Santos:19-c}.

From the above discussion emerges an important element for quantum batteries called \textit{passive states}. Passive states were originally introduced in Ref.~\cite{Pusz:78}, with recent reviews and a detailed discussion on its relation with the Eq.~\eqref{a1} given in Refs.~\cite{Allahverdyan:04,Perarnau-Llobe:15}. In summary, passive states are associated with a state where no amount of work can be extracted from the QB in a cyclic unitary process, so that the uniqueness of passive states are not possible in general~\cite{Perarnau-Llobe:15}. Then, consider a reference Hamiltonian $H$ and a given state $\rho$, we can write both of them in their respective eigenbasis (increasing for $H$, decreasing for $\rho$)
\begin{align}
H&=\sum_{j}\varepsilon_{j}\vert\varepsilon_{j}\rangle\langle\varepsilon_{j}\vert, ~~ \varepsilon_{j+1}\geqslant\varepsilon_{j} ~~ \forall ~ j \text{ , } \\
\rho &=\sum_{j}r_{j}\vert r_{j}\rangle\langle r_{j}\vert, ~~  r_{j+1}\leqslant r_{j} ~~ \forall ~ j \text{ . }
\end{align}

Then, we say that the state $\rho$ is passive with respect to $H$ if and only if (i) $\rho$ and $H$ are diagonal in the same basis being valid the commutation relation $[\rho,H]\!=\!0$ and (ii) $\rho$ contains no population inversion, i.e., $\varepsilon_{i}\!<\!\varepsilon_{j} \Rightarrow r_{i} \!\geqslant\! r_{j}$. We have denoted passive states by $\sigma_{\rho}$, thus it takes the following form~\cite{Allahverdyan:04,Perarnau-Llobe:15}
\begin{align}
\sigma_{\rho}=\sum_{j}r_{j}\vert\varepsilon_{j}\rangle\langle\varepsilon_{j}\vert \text{ . }
\end{align}

The understanding of passive state is important to define the empty battery state, since no work can be extracted from a passive state we therefore say the battery is empty when the battery state is passive. Intuitively, non-passive states are called \textit{active states} because energy can be extracted from the QB~\cite{Binder:15}. In this sense, it is evident that the charging process of a battery does not mean increase the system energy, it means we need to increase its ergotropy. Of course, in some cases we can have simultaneously both increasing ergotropy and system energy. However, it is not the case here because the charging (discharging) processes are not necessarily unitary and the quantum system may interact with its surrounding. For this reason, it is important to define the ratio between the extractable work and the energy value of the battery as
\begin{align}
R (t) =\frac{\mathcal{W} (t)}{\Delta E (t)},
\end{align}
with $\mathcal{W} (t)$ and $\Delta E (t) = \tr [H\rho(t)] - \tr[H\rho(0)]$ being the instantaneous ergotropy and energy variation, respectively. In addition, its instantaneous charging power is defined here from available work in the battery as
\begin{align}
\Pcal (t)= \lim_{\Delta t \rightarrow 0 }\frac{\mathcal{W}(t+\Delta t) - \mathcal{W}(t)}{\Delta t} = \frac{d \mathcal{W}(t)}{dt} \text{ . }
\end{align}
In fact, how fast can a battery charged (or discharged), it depends on its power. In the next section, we introduce the model of open QB used here by assuming that its performance may be affected by a dissipative environment.

\begin{figure}
	\centering
	\includegraphics[scale=0.45]{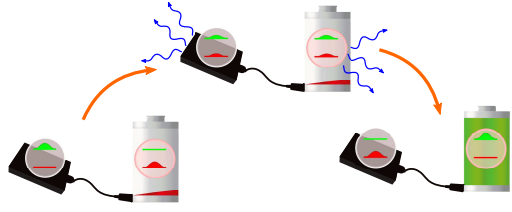}
	\caption{Sketch of the system considered here, where the charger and the battery are considered as two-level systems. Initially the system is composed of the charger with maximum ergotropy and it will transfer this energy to the battery. Along the evolution losses are considered and it can affect the charging performance of the battery. In the ideal case, all energy is transferred and the battery ergotropy is maximum.}\label{Scheme}
\end{figure}

\section{The model}\label{sec3}

Here, we consider a model consisting of four components: two two-level systems (two qubits) interacting with each other at a given time interval, such that one is considered as a quantum battery $B$ and another as charger $A$, and each qubit is locally contacted with an independent amplitude damping reservoir $E$ (see Fig.\ref{Scheme}). So, the total Hamiltonian is given by
\begin{align}
H=H_{0}+H_{\text{int}},
\end{align}
where
\begin{align}
H_{0}= \frac{\omega_{0}}{2}\sigma^{A}_{z}+\frac{\omega_{0}}{2}\sigma^{B}_{z}+\sum_{k}\omega^{A}_{k}a^{\dagger}_{k}a_{k}+\sum_{k}\omega^{B}_{k}b^{\dagger}_{k}b_{k},
\end{align}
in which the two first terms indicate the free Hamiltonians of the charger and the QB, respectively, as well as two last terms denote the free Hamiltonians of the independent environments coupled to the charger and the battery, respectively. In our system $\omega_{0}$ is the transition frequency of the battery and charger, where $\sigma^{j}_{z}$ $(j\!=\!A,B)$ represent the Pauli operator in $z$ direction of the systems $A$ and $B$, respectively. Also,
$\omega^{j}_{k}$ $(j\!=\!A,B)$ is the frequency of the $k$th mode of the environmen $A$ and $B$, respectively,
$a^{\dagger}_{k}$ and $ b^{\dagger}_{k}$  $(a_{k},b_{k})$ are the creation (annihilation) operators
corresponding to the $k$-th mode of the environments.

To study the energy transfer from the charger to the battery, we consider the interaction Hamiltonian defined as
\begin{align}
H_{\text{int}}=\kappa(\sigma^{A}_{+}\sigma^{B}_{-}+\sigma^{A}_{-}\sigma^{B}_{+})+\sum_{k}(g^{A}_{k}\sigma^{A}_{+}a_{k}+g^{\ast A}_{k}\sigma^{A}_{-}a^{\dagger}_{k}) + \sum_{k}(g^{B}_{k}\sigma^{B}_{+}b_{k}+g^{\ast B}_{k}\sigma^{B}_{-}b^{\dagger}_{k}),
\end{align}
where the first term describes the interaction between the two qubits with the coupling constant $\kappa$, and
two last summations show the coupling between the charger and the battery to their respective environments, that $g^{j}_{k}$ $(j\!=\!A,B)$ is the coupling constant between the $j$-th qubit and the $k$-th mode of its corresponding environment. Here, we are denoting $\sigma^{j}_{\pm}=(\sigma^{j}_{x}\pm\sigma^{j}_{y})/2$ as the raising and lowering Pauli operators of the $j$-th qubit.

In this work, we exactly obtain the time evolution of the total system. For this purpose, we work in the interaction picture defined by
\begin{align}
H^{\text{I}}_{\text{int}}(t)=e^{i H_{0}t}~H_{\text{int}}~e^{-i H_{0}t} \text{ , }
\end{align}
in which the Hamiltonian takes the form
\begin{align}
H^{\text{I}}_{\text{int}}(t)= \kappa(\sigma^{A}_{+}\sigma^{B}_{-}+\sigma^{A}_{-}\sigma^{B}_{+}) +H^{\text{I}}_{1}  \text{ , }
\end{align}
where
\begin{align}
H^{\text{I}}_{1}=\sum_{k}(g^{A}_{k}\sigma^{A}_{+}a_{k}e^{i(\omega_{0}-\omega^{A}_{k})t} + \text{h.c.})+\sum_{k}(g^{B}_{k}\sigma^{B}_{+}b_{k}e^{i(\omega_{0}-\omega^{B}_{k})t}+ \text{h.c.}) \text{ . }
\end{align}

As shown in~\ref{SolAp}, the total system dynamics can be written as a superposition given by
\begin{equation}
\vert\psi(t)\rangle =\left[\mu(t)\vert e , g\rangle +\nu(t)\vert g , e\rangle\right]_{\text{S}}\otimes\vert 0,0\rangle_{\text{B}} + \vert g , g\rangle_{\text{S}}\otimes\left(\sum_{k}\eta^{\text{A}}_{k}(t)\vert 1_{k} , 0 \rangle+ \eta^{\text{B}}_{k}(t)\vert 0 , 1_{k}\rangle\right)_{\text{R}}, \label{Sol}
\end{equation}
with $\mu(t)$, $\nu(t)$, $\eta^{\text{B}}_{k}(t)$ and $\eta^{\text{A}}_{k}(t)$ being functions to be determined. We are using the notation $\vert \psi , \phi \rangle_{\text{S}} = \vert \psi\rangle_{\text{A}}\vert \phi\rangle_{\text{B}}$ for the system and $\vert n,m \rangle_{\text{R}} = \vert n^{A}\rangle\vert m^{B}\rangle$ for the Fock basis of the reservoir.

Therefore, we analytically solve the system dynamics under the initial condition $\ket{\psi(0)} = \ket{e , g}_{\text{S}} \otimes \vert 0,0\rangle_{\text{R}}$, a situation in which the QB is empty and the charger has maximal energy, the dynamical equation for $\mu(t)$ and $\nu(t)$ are obtained from respective inverse Laplace transform of (see~\ref{SolAp})
	\begin{align}
	\tilde{\mu}(s)=\frac{s+\tilde{k}(s)}{[s+\tilde{k}(s)]^2+\kappa^{2}} \text{ , } ~~
	\tilde{\nu}(s)=-\frac{i\kappa}{[s+\tilde{k}(s)]^2+\kappa^{2}} \text{ , } \label{b7}
	\end{align}
with $\tilde{k}(s)$ being the Laplace transform of $k(t)$, with $k(t-t^{\prime})$ being the correlation function given by $k(t-t^{\prime})=\frac{1}{2}\gamma\lambda e^{(-\lambda+ i \Delta) \vert t-t^{\prime}\vert}$~\cite{Breuer:12}, where the parameter $\gamma$ is an effective coupling constant which is connected to the relaxation time of the qubit $\tau_{\text{R}} \approx\gamma^{-1}$, $\lambda$ is the width of the Lorentzian spectrum that is linked to the environment correlation time $\tau_{\text{E}}\approx\lambda^{-1}$ and $ \Delta $ is the detuning of $\omega_{0}$ and the central frequency of the environment.

The ergotropy and instantaneous energy of the QB and the charger is computed from reduced density matrix for each system, then according to Eq. (\ref{Sol}), the reduced density matrix of the QB and the charger are respectively as follows
\begin{subequations}
	\begin{align}
	\rho_{\text{B}}(t) &= \vert\nu(t)\vert^{2} \ket{e}\bra{e}_{\text{B}} +  \left[1-\vert\nu(t)\vert^{2}\right] \ket{g}\bra{g}_{\text{B}}  \text{ , } \label{QBdensityMatrix} \\
	\rho_{\text{A}}(t) &= \vert\mu(t)\vert^{2} \ket{e}\bra{e}_{\text{A}} +  \left[1-\vert\mu(t)\vert^{2}\right] \ket{g}\bra{g}_{\text{A}}  \text{ , }
	\end{align}
\end{subequations}
so that it is straightforward to show that the internal energy of the battery and the charger can be obtained as, respectively,
\begin{align}\label{b9}
E_{\text{B}}(t)=\hbar \omega_{0}\vert\nu(t)\vert^{2} \text{ , } ~~ E_{\text{A}}(t)= \hbar \omega_{0}\vert\mu(t)\vert^{2} \text{ . }
\end{align}

As previously mentioned, the above equations do not represent the available energy that can be extracted from the QB after the charging process. To find the ergotropy of the QB we can use the Eq.~\eqref{QBdensityMatrix}, where it is possible to see that for $\vert\nu(t)\vert^{2} \leq 1/2$, no amount of energy can be extracted from the QB by unitary processes so that $\mathcal{W}_{\text{B}}(t) = 0$ for all $t$ with $\vert\nu(t)\vert^{2} \leq 1/2$. However, when $\vert\nu(t)\vert^{2} > 1/2$, then we have available energy because the ergotropy is given by $\mathcal{W}_{\text{B}}(t)|_{\vert\nu(t)\vert^{2} > 1/2} = \hbar \omega_{0}\left(2\vert\nu(t)\vert^{2}-1\right)$. Therefore, by using the Heaviside function $\Theta(x-x_{0})$, which satisfies $\Theta(x-x_{0}) = 0$ for $x<x_{0}$, $\Theta(x-x_{0}) = 1/2$ for $x=x_{0}$ and $\Theta(x-x_{0}) = 1$ for $x>x_{0}$, we can write the
\begin{equation}\label{b10}
\mathcal{W}_{\text{B}}(t)= \Theta\left( \vert\nu(t)\vert^{2} -1/2\right) \mathcal{W}_{\text{B}}(t)|_{\vert\nu(t)\vert^{2} > 1/2}.
\end{equation}

In the following, by using the above equation we can study the charger-mediated charging process of the battery under Markovian and non-Markovian dynamics.

\section{Markovian and non-Markovian effects on battery charging}

Now we investigate the amount of available energy transferred from the charger to the QB after the charging process under Markovian and non-Markovian effects. As a first assumption, in our calculations we consider the case where both charger and battery are in resonance with the reservoir modes by putting $\Delta\!=\!0$. As we shall see, it would be worth to consider $\Delta\!\neq\!0$, but $\Delta\!\neq\!0$ means a non-maximum effective coupling between the system and the reservoir. For this reason, we consider the case where we have maximum coupling in order to provide a good approach to the charging and discharging process in the worst decohering scenario. Moreover, in order to distinguish the strong coupling regime from the weak coupling regime, we can define the dimensionless positive parameter $R=\gamma/\lambda$. It has been demonstrated that for $R\ll 1$ the dynamics is divisible (Markovian), while for $R\gg 1$ it becomes non-divisible (non-Markovian)~\cite{Luo:12,Breuer:09,Wolf:08,Rivas:10,Garcapintos:19,Obando:15,Passos:19-b,Paula:16,Lorenzo:13,Breuer:12}. It is important to emphasize the ratio between $\gamma$ and $\kappa$ plays a significant role in energy transfer. Hence, we display two distinct regimes where we have (i) an underdamped regime occurring for $\kappa \gg \gamma$, that means the coupling between the QB and the charger is stronger than the interaction between them and their corresponding environments and, conversely, (ii) the overdamped regime obtained for $\kappa\ll\gamma$. As a third situation, we also consider (ii) an intermediate coupling regime with $\kappa \sim \gamma$ (for example, $\kappa\!=\!\gamma$).

\begin{figure}[t]
    \centering
     \includegraphics[scale=0.45]{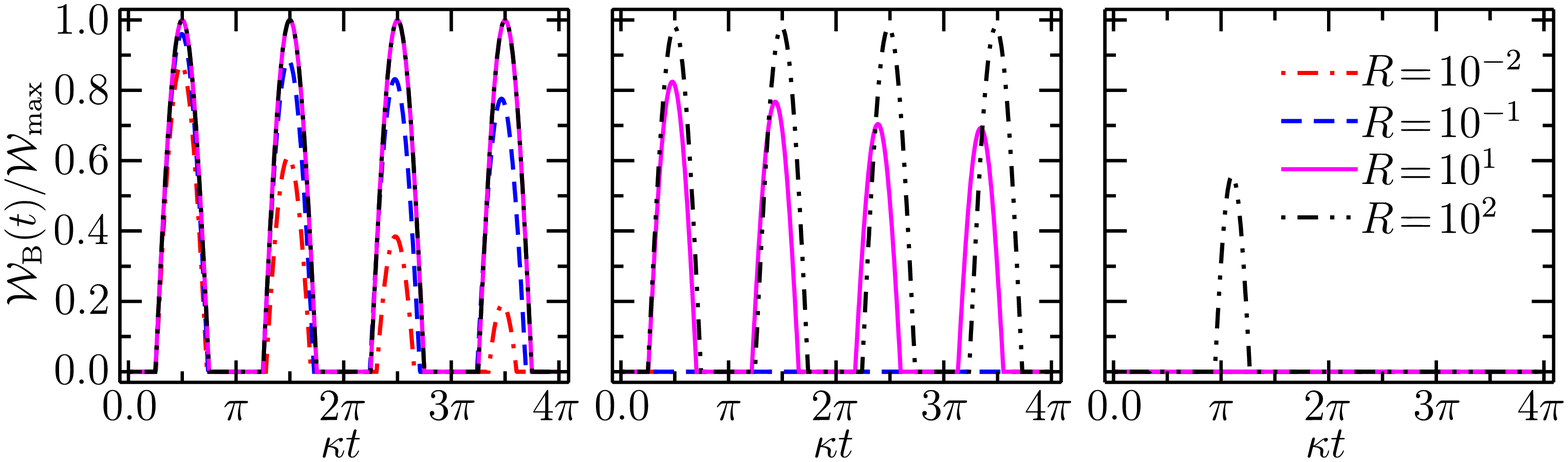}
       \caption{Ergotropy $\mathcal{W}_{\text{B}}(t)$ (in multiple of $\mathcal{W}_{\text{max}}$) as a function of the dimensionless quantity $\kappa t$ for different values of the parameter $R$. We consider effects of the coupling strength for (left) underdamped regime with $\gamma\!=\!0.05\kappa$, (middle) intermediate regime with $\gamma\!=\!\kappa$ and (right) overdamped case where $\gamma\!=\!10\kappa$. Numerically we can show that in case where $\gamma\!>\!100$ we get $\mathcal{W}_{\text{B}}(t)\!=\!0$ for all $t\in[0,4\pi]$.}\label{Ergo1}
\end{figure}

By driving the system (QB and charger) without interaction with the baths, one finds the required time $\tau_{\text{ch}}$ to fully transfer the energy from charger to QB and such parameter works as a natural time-scale for the charging process. By definition, the time $\tau_{\text{ch}}$ is obtained from $\mathcal{W}_{\text{B}}(\tau_{\text{ch}})\!=\!\mathcal{W}_{\text{max}}$, where $\mathcal{W}_{\text{max}} = \hbar \omega$ is the maximum ergotropy that can be stored in the QB. Since the excitation present in the charger will be transferred to the QB in absence of surroundings interacting with the system, we can solve the Schrödinger equation and find that the probability of transferring the excitation from the charger to battery is
\begin{align}\label{probability}
p(t) = |\interpro{g,e}{\psi(t)}|^2 = |\bra{g,e}e^{-\frac{i}{\hbar}\kappa(\sigma^{A}_{+}\sigma^{B}_{-}+\sigma^{A}_{-}\sigma^{B}_{+})t}\vert\psi(0)\rangle|^2 = \sin^2(\kappa t) \text{ , }
\end{align}
so that $p(t)\!=\!1$ when $\tau_{\text{ch}}\!=\!\pi/2\kappa$. In this case, it is possible to define the average power to energy transfer given by
\begin{align}
\Pcal_{\text{av}} = \frac{\mathcal{W}_{\text{B}}(t)-\mathcal{W}_{\text{B}}(t_{0})}{t - t_{0}} = \frac{\hbar \omega_{0}}{\tau_{\text{ch}}} = \frac{2 \hbar \kappa \omega_{0}}{\pi} \text{ , } \label{MaxP}
\end{align}
for a non-decohering process. Now, when we consider the decohering effects on the system the above quantities can be drastically changed, as shown in Fig.~\ref{Ergo1}.

In Fig.~\ref{Ergo1}, we consider the coupling regimes from underdamped (left-hand graph) to the overdamped case (right-hand graph) for different values of $R$. In the underdamped regime it is possible to see that the smaller the parameter $R$ ($R\!=\!0.01$ and $R\!=\!0.1$), the worse will be the charging performance of the battery. On the other hand, by increasing the parameter $R$ ($R\!=\!10$ and $R\!=\!100$) we can achieve the regime of maximum power charging of the battery given in Eq.~\eqref{MaxP}, since the first maximum peak of the dashed-dot-dot black and the magenta curve happens for $t\!=\!\tau_{\text{ch}}$. More precisely, at $t\!=\!\tau_{\text{ch}}$ we have $p(\tau_{\text{ch}})\!\approx\!99.995\%$ for both cases $R\!=\!10$ and $R\!=\!100$, so that it means we transfer $99.995\%$ of the maximum energy $(\hbar \omega_{0}$) to the battery. Therefore, the non-Markovian dynamics of the system allows for an optimal energy transfer in the underdamped coupling regime. In order to get a direct comparison with the Markovian case, at instant $t\!=\!\tau_{\text{ch}}$ we have $p(\tau_{\text{ch}})\!\approx\!87.057\%$ and $p(\tau_{\text{ch}})\approx 95.948\%$ for $R\!=\!0.01$ and $R\!=\!0.1$, respectively. More drastically, for the cases where we have intermediate ($\gamma\!=\!\kappa$) and overdamped ($\gamma\!=\!10\kappa$) coupling regimes, the battery is not charged for the cases $R = 0.01$ and $R = 0.1$, with the best charging process obtained for $R\!=\!100$.

\begin{figure}
	\centering
	\includegraphics[scale=0.43]{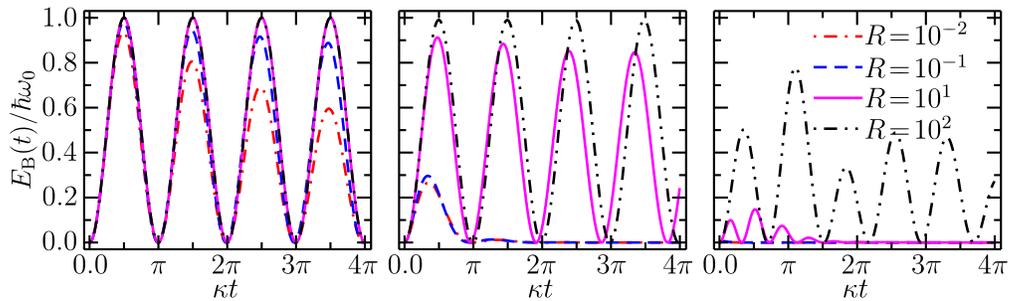}
	\caption{Instantaneous energy $E_{\text{B}}(t)$ in the battery (as multiple of $\hbar\omega_{0}$) as a function of the dimensionless quantity $\kappa t$ for different values of the parameter $R$. We consider effects of the coupling strength for (left) underdamped regime with $\gamma\!=\!0.05\kappa$, (middle) intermediate regime with $\gamma\!=\!\kappa$ and (right) overdamped case where $\gamma\!=\!10\kappa$.}\label{Energy}
\end{figure}

It is worth mentioning here that the value zero and constant for some time intervals $\Delta t$ does not mean the system energy is zero. The energy in the QB is computed from the quantum mechanics postulates as $E_{\text{B}}(t)\!=\!\trs{H_{0}\rho^{\text{B}}(t)} - E_{\text{gs}}$, where $E_{\text{gs}}\!=\!\hbar\omega_{0}/2$ is the ground state energy for our battery and $\rho^{\text{B}}(t)$ is the battery reduced density matrix. As we can see from Fig.~\ref{Energy}, the energy flux in the charger-battery system is continuum along the system dynamics. However, such energy may be stored in the battery as a non-extractable amount of energy. In fact, for example, if the charger-battery system state is maximally entangled $\ket{\psi_{\text{bell}}}\!=\!(\ket{e,g}+\ket{g,e})/\sqrt{2}$, we have some energy in the battery but it can not be extracted from the battery by unitary processes because its reduced density matrix reads $\rho_{\psi_{\text{bell}}}^{\text{B}}\!=\!\1$. This result allows us to clearly see the main difference between energy of the system and system ergotropy.

\section{Markovian and non-Markovian self-discharging of quantum batteries}

In classical batteries we have a phenomenon known as \textit{self-discharging} of batteries associated with undesired effects that compromise the storing performance of the battery~\cite{Wu:20,Galushkin:12,Wenhua:14,Shinyama:06}. The self-discharging process is characterized by a loss of charge in the battery, even when the battery is not coupled to some consumption hub. In the quantum realm, the self-discharging mechanism has been introduced as a consequence of the inevitable coupling of the QB with some external reservoir~\cite{Santos:19-a}. Here we are interested in investigating the role of Markovian and non-Markovian in the self-discharging of a two-level quantum battery, since such mechanism has been considered in a tree-level QB~\cite{Santos:19-a}.

\begin{figure}[t!]
	\centering
	\includegraphics[scale=0.43]{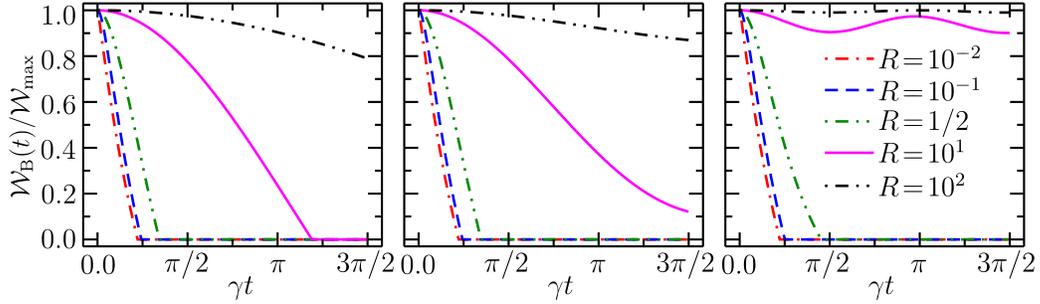}
	\caption{Ergotropy $\mathcal{W}_{\text{B}}(t)$ in the battery (as multiple of $\mathcal{W}_{\text{max}}$) in the function of the dimensionless quantity $\gamma t$ for different values of the parameter $R$. Here we consider the influence of $\Delta$ by choosing (left) $\Delta\!=\!0$, (middle) $\Delta\!=\!0.5\gamma$ and (right) $\Delta\!=\!2\gamma$.}\label{SD}
\end{figure}

To this end, first we consider the system as composed by the QB and its reservoir. This system can be considered by taking the limit $\kappa \rightarrow 0$ and starting the QB at state $\ket{e}_{\text{B}}$ and the reservoir in state $\ket{0}_{\text{R}}$. Under these assumptions, the evolved state of the system QB-reservoir is described by
\begin{align}
\ket{\psi_{\text{sd}}(t)} = \nu_{\text{sd}}(t)\ket{e}_{\text{B}}\otimes\ket{0}_{\text{R}} +
\sum_{k}\eta^{\text{sd}}_{k}(t)\vert g\rangle_{\text{B}} \otimes \ket{1_{k}}_{\text{R}}
\end{align}
where the subscript ``sd" means self-discharging. Again, it is possible to show that the reduced density matrix of the QB is given by the Eq.~\eqref{QBdensityMatrix} but now the coefficient is obtained from the inverse Laplace transform of
\begin{align}
\tilde{\nu}_{\text{sd}}(s) = \frac{1}{s+\tilde{k}(s)} \text{ , }
\end{align}
which can be obtained from Eq.~\eqref{b7Ap} of the~\ref{SolAp} with $\kappa \rightarrow 0$. Then, the solution to the above equation provides
\begin{equation}
|\nu_{\text{sd}}(t)| = e^{-\frac{t \lambda }{2}}\left\vert \cosh \left( \frac{1}{2} t \xi\right) +
\frac{\lambda - i\Delta}{\xi}\sinh \left( \frac{1}{2} t \xi\right)
\right\vert \text{ , }
\end{equation}
where $\xi^2 = (\lambda - i\Delta)^{2} - 2\gamma \lambda$. As a first result, let us consider the case where $\Delta \rightarrow 0$, so that the above equation becomes
\begin{equation}
\lim_{\Delta \rightarrow 0}|\nu_{\text{sd}}(t)| = e^{-\frac{t\gamma }{2R}}\left\vert \cosh \left( \frac{t\gamma \sqrt{1 - 2R}}{2R}\right) + \frac{1}{\sqrt{1 - 2R}} \sinh \left( \frac{t\gamma \sqrt{1 - 2R}}{2R}\right)
\right\vert \text{ , }
\end{equation}
where we already used $R\!=\!\gamma/\lambda$. Now, notice that the parameter $R$ allows us to provide different self-discharging processes, so that it is evident that such process depends on the Markovian and non-Markovian regime. Due to the quantity $1 - 2R$, here we can highlight three important regimes (i) $R\!<\!1/2$, (ii) $R\!=\!1/2$ and (i) $R\!>\!1/2$. Therefore, in Fig.~\ref{SD} we present the ergotropy as for different values of $R$, including the case $R\!=\!1/2$. However, regardless of the regime we consider the battery becomes fully discharged in limit $t\gamma \rightarrow \infty$, then the question here is how this discharging process happens.

By comparing the curves in graphs of the Fig.~\ref{SD}, as already mentioned, by increasing the value of
 $\Delta \neq 0$ we decrease the effective coupling of the system with the reservoir and the self-discharging becomes slower than in the cases where $\Delta = 0$. As for the role of the Markovianity and non-Markovianity, in summary the self-discharging of the QB follows the same behavior as the charging process, where the bigger $R$, the better the performance of the QB performance against its self-discharging process. Therefore, in this case, the non-Markovian processes are preferable regarding Markovian ones.

\section{Conclusion}

In this paper, we provided a model for robust quantum batteries, inspired by the inevitable interaction of quantum systems with their environments. In the charging process the quantum battery interacts with the charger, while each of them is in contact with independent environments. Our results show that the memory effects due to non-Markovian dynamics are beneficial for the charging process of our battery, where we can approximately achieve the maximum charge of the battery. Therefore, this energy can be later used and transferred to consumption hubs. Moreover, we studied the phenomenon related to the self-discharging of the battery. Such study is motivated by its classical counterpart, where these undesired effects limit the performance of classical and quantum batteries in storing energy~\cite{Santos:19-a}. We showed that the self-discharge time of the battery in non-Markovian dynamics is longer than the Markovian dynamics, as a consequence the stored energy can be preserved for long time in the presence of non-Markovian dynamics. Our results suggest that memory effects can play an important role in improving the performance of quantum batteries in the framework of the open system approach.

\section*{Acknowledgments}

This work has been supported by the University of Kurdistan. F. T. Tabesh and S. Salimi thank Vice Chancellorship of Research and Technology,  University of Kurdistan. A. C. Santos acknowledges the financial support through the research grant from the São Paulo Research Foundation (FAPESP) (grant 2019/22685-1).

\appendix

\section{Solution for dynamics}\label{SolAp}

To solve the dynamics of the system considered here we use the fact that the number operator is a constant of motion i.e; $[H,N]=0$, where
\begin{align}
N=\sum_{j=A,B}\sigma^{j}_{+}\sigma^{j}_{-}+\sum_{k}a^{\dagger}_{k}a_{k}+\sum_{k}b^{\dagger}_{k}b_{k}  \text{ . }
\end{align}

Therefore, if we assume that the initial state of the system is a combination of the ground ($\vert g\rangle$) and excited ($\vert e\rangle$) states of the battery and  the charger such that the total initial state given by ($\vert\mu(0)\vert^{2}+\vert\nu(0)\vert^{2}=1$)
\begin{align}\label{eq16}
\vert\psi(0)\rangle=[\mu(0)\vert e\rangle_{\text{A}}\vert g\rangle_{\text{B}}+\nu(0)\vert g\rangle_{\text{A}}\vert e\rangle_{\text{B}}]\otimes\vert 0^{A}\rangle\vert 0^{B}\rangle,
\end{align}
with $\vert \psi\rangle_{\text{A}}$ and $\vert \phi\rangle_{\text{B}}$ denoting the charger and the battery states, and $\vert 0^{j}\rangle$ being the vacuum state of the corresponding environment, then we can write the general dynamics of the system in basis spanned by a single excitation subspace as
\begin{align}\label{eq17}
	\vert\psi(t)\rangle =&[\mu(t)\vert e , g\rangle +\nu(t)\vert g , e\rangle]_{\text{S}}\otimes\vert 0,0\rangle_{\text{B}} + \vert g , g\rangle_{\text{S}}\otimes\left[\sum_{k}\eta^{A}_{k}(t)\vert 1_{k} , 0 \rangle+ \eta^{B}_{k}(t)\vert 0 , 1_{k}\rangle\right]_{\text{R}},
	\end{align}
in which from now on we are using the notation $\vert \psi , \phi \rangle_{\text{S}} = \vert \psi\rangle\vert \phi\rangle$ for the system and $\vert n,m \rangle_{\text{R}} = \vert n^{A}\rangle\vert m^{B}\rangle$ for the Fock basis of the bath. Here $\vert 1_{k}\rangle$ characterizes a state of the corresponding environment with one excitation in the $k$-th mode. Now, by putting the above \textit{ansatz} in Schrödinger equation one can find the following differential equations for the probability coefficients $\mu(t)$, $\nu(t)$ and $\eta^{j}_{k}(t)$
\begin{subequations}\label{b1}
\begin{align}
\dot{\mu}(t)&=-i(\kappa \nu(t)+\sum_{k}\eta^{A}_{k}(t)g^{A}_{k}e^{i(\omega_{0}-\omega^{A}_{k})t})  \text{ , } \\
\dot{\nu}(t)&=-i(\kappa \mu(t)+\sum_{k}\eta^{B}_{k}(t)g^{B}_{k}e^{i(\omega_{0}-\omega^{B}_{k})t}) \text{ , }  \\
\dot{\eta}^{A}_{k}(t)&=-i\mu(t)g^{A\ast}_{k}e^{-i(\omega_{0}-\omega^{A}_{k})t} \text{ , }  \\
\dot{\eta}^{B}_{k}(t)&=-i\nu(t)g^{B\ast}_{k}e^{-i(\omega_{0}-\omega^{B}_{k})t}  \text{ . }
\end{align}
\end{subequations}

Now, by using the two last differential equations in Eq.~(\ref{b1}) with the initial conditions $\eta^{A}_{k}(0)=\eta^{B}_{k}(0)=0$, the probability coefficients $\eta^{j}_{k}(t)$ can be written as
\begin{subequations}
\begin{align}
\eta^{A}_{k}(t)&=-ig^{A\ast}_{k}\int^{t}_{0}\mu(t^{\prime})e^{-i(\omega_{0}-\omega^{A}_{k})t^{\prime}}dt^{\prime}\text{ , } \label{b2} \\ \eta^{B}_{k}(t)&=-ig^{B\ast}_{k}\int^{t}_{0}\nu(t^{\prime})e^{-i(\omega_{0}-\omega^{B}_{k})t^{\prime}}dt^{\prime}\text{ . }\label{b3}
\end{align}
\end{subequations}

By applying above equations, the differential equations for $\mu(t)$ and $\nu(t)$ can be obtained as
\begin{subequations}\label{b5}
\begin{align}
\dot{\mu}(t)&=-i\kappa \nu(t)-\int^{t}_{0}k_{A}(t-t^{\prime})\mu(t^{\prime})dt^{\prime} \text{ , } \\
\dot{\nu}(t)&=-i\kappa \mu(t)-\int^{t}_{0}k_{B}(t-t^{\prime})\nu(t^{\prime})dt^{\prime} \text{ , }
\end{align}
\end{subequations}
where the kernels $k_{A}(t-t^{\prime})$ and $k_{B}(t-t^{\prime})$ are given by
\begin{align}
k_{A}(t-t^{\prime})=\sum_{k}\vert g^{A}_{k}\vert^{2}e^{i(\omega_{0}-\omega^{A}_{k})(t-t^{\prime})} \text{ , } ~ k_{B}(t-t^{\prime})=\sum_{k}\vert g^{B}_{k}\vert^{2}e^{i(\omega_{0}-\omega^{B}_{k})(t-t^{\prime})} \text{ . }
\end{align}

For simplicity, we suppose $k_{A}(t-t^{\prime})=k_{B}(t-t^{\prime})=k(t-t^{\prime})$. In continuum limit, the correlation function $k(t-t^{\prime})$ can be expressed as
\begin{align}
k(t-t^{\prime})=\int^{\infty}_{0}d\omega J(\omega)e^{i(\omega_{0}-\omega)(t-t^{\prime})},
\end{align}
in which $J(\omega)$ is the spectral density of the environments. In the model considered in our study, it is taken as
\begin{align}
J(\omega)=\frac{\gamma ~ \lambda^{2}}{2\pi[(\omega_{0}-\omega-\Delta)^{2}+\lambda^{2}]},
\end{align}
where $\gamma$ is an effective coupling constant which is connected to the relaxation time of the qubit $\tau_{R} \approx\gamma^{-1}$ and $\lambda$ is the width of the Lorentzian spectrum that is linked to the environment correlation time $\tau_{E}\approx\lambda^{-1}$ and $ \Delta $ is the detuning of $\omega_{0}$ and the central frequency of the environment. By some calculations, one can obtain the correlation function as $k(t-t^{\prime})=\frac{1}{2}\gamma\lambda e^{(-\lambda+ i \Delta) \vert t-t^{\prime}\vert}$~\cite{Breuer:12}.

At this step, let us define
\begin{align}
z(t) &=\mu(t)+\nu(t) \text{ , } ~~ w(t) =\mu(t)-\nu(t).
\end{align}
Then, by regarding Eq. (\ref{b5}), one can write the following equations
\begin{subequations}
	\begin{align}
	\dot{z}(t)&=-i\kappa z(t)-\int^{t}_{0}k(t-t^{\prime})z(t^{\prime})dt^{\prime} \text{ , } \\
	\dot{w}(t)&=i\kappa w(t)-\int^{t}_{0}k(t-t^{\prime})w(t^{\prime})dt^{\prime} \text{ . }
	\end{align}
\end{subequations}

Applying the Laplace transform, the above equations becomes
\begin{subequations}\label{b11}
	\begin{align}
	s\tilde{z}(s)-z(0)&=-i\kappa \tilde{z}(s)-\tilde{k}(s)\tilde{z}(s) \text{ , }\\
	s\tilde{w}(s)-w(0)&=i\kappa \tilde{w}(s)-\tilde{k}(s)\tilde{w}(s)\text{ , }
	\end{align}
\end{subequations}
in which $\tilde{z}(s)$, $\tilde{w}(s)$ and $\tilde{k}(s)$ are the Laplace transforms of $z(t)$, $w(t)$ and $k(t-t^{\prime})$, respectively. To solve the dynamics, we begin with first equation of (\ref{b11}) in the main text
\begin{align}\label{b14}
s\tilde{z}(s)-z(0)=-i\kappa ~ \tilde{z}(s)-\tilde{k}(s)\tilde{z}(s) \text{ , }
\end{align}
and replace $s$ with $s $, then we have
\begin{align}\label{b12}
\tilde{z}(s )[s  +\tilde{k}(s )]=z(0)-i\kappa ~\tilde{z}(s) \text{ , }
\end{align}
now, exchange  $\delta$ by $-\delta$ in Eq.(\ref{b12}), the above equation takes the following form
\begin{align}\label{b13}
\tilde{z}(s )=\frac{z(0)-i\kappa ~\tilde{z}(s)}{[s +\tilde{k}(s )]} \text{ . }
\end{align}
Substituting Eq.(\ref{b13})  into (\ref{b14}), we obtain
\begin{align}
\tilde{z}(s)=z(0)(\frac{1}{s+\tilde{k}(s)+\frac{\kappa^{2}}{Q(s)}}-\frac{i\kappa}{Q(s)[s+\tilde{k}(s)]+\kappa^{2}}) \text{ , }
\end{align}
where $Q(s)=s +\tilde{k}(s )$. Similarly, starting from the second equation in (\ref{b11}), one can get the following equation for $\tilde{w}(s)$
\begin{align}
\tilde{w}(s)=w(0)(\frac{1}{s+\tilde{k}(s)+\frac{\kappa^{2}}{Q(s)}}+\frac{i\kappa}{Q(s)[s+\tilde{k}(s)]+\kappa^{2}}) \text{ . }
\end{align}
On the other hand, by following definitions
\begin{align}
\tilde{\mu}(s)&=\frac{1}{2}[\tilde{z}(s)+\tilde{w}(s)] \text{ , } ~~ \tilde{\nu}(s)=\frac{1}{2}[\tilde{z}(s)-\tilde{w}(s)] \text{ , }
\end{align}
and some straightforward calculations, they can be written as
\begin{subequations}
	\begin{align}
	\tilde{\mu}(s)&=\frac{[s +\tilde{k}(s )]\mu(0)-i\kappa\nu(0)}{[s+\tilde{k}(s)][s +\tilde{k}(s )]+\kappa^{2}} \text{ , } \label{b6Ap} \\
	\tilde{\nu}(s)&=\frac{[s +\tilde{k}(s )]\nu(0)-i\kappa\mu(0)}{[s+\tilde{k}(s)][s +\tilde{k}(s )]+\kappa^{2}} \text{ . } \label{b7Ap}
	\end{align}
\end{subequations}

\end{document}